\documentclass{article}

\usepackage{amssymb,amsfonts,amsmath}
\usepackage{cite,enumerate,float,indentfirst}
\usepackage{color}

\def\be{\begin{eqnarray}}
\def\ee{\end{eqnarray}}
\def\nn{\nonumber}

\def\p{\partial}

\def\Tr{{\rm Tr}\,}

\def\l[{\phantom.[}

\definecolor{red}{rgb}{1,0,0}
\definecolor{orange}{rgb}{1,0.5,0}
\definecolor{violet}{rgb}{0.7,0,1}

\hoffset= - 1.0in         

\begin{document}

\title{\vspace{.1cm}{\Large {\bf On the complete perturbative solution of one-matrix  models}\vspace{.2cm}}
\author{
{\bf A.Mironov$^{a,b,c}$}\footnote{mironov@lpi.ru; mironov@itep.ru}\ \ and
\ {\bf A.Morozov$^{b,c}$}\thanks{morozov@itep.ru}}
\date{ }
}

\maketitle

\vspace{-5.5cm}

\begin{center}
\hfill FIAN/TD-08/17\\
\hfill IITP/TH-07/17\\
\hfill ITEP/TH-13/17
\end{center}

\vspace{3.3cm}

\begin{center}
$^a$ {\small {\it Lebedev Physics Institute, Moscow 119991, Russia}}\\
$^b$ {\small {\it ITEP, Moscow 117218, Russia}}\\
$^c$ {\small {\it Institute for Information Transmission Problems, Moscow 127994, Russia}}

\end{center}

\vspace{.5cm}

\begin{abstract}
We summarize the recent results about complete solvability of
Hermitian and rectangular complex matrix models.
Partition functions have very simple character expansions with
coefficients made from dimensions of representation of the linear
group $GL(N)$, and arbitrary correlators in the Gaussian phase
are given by finite sums over Young diagrams of a given size,
which involve also the well known characters of symmetric group.
The previously known integrability and Virasoro constraints are
simple corollaries, but no vice versa:
complete solvability is a peculiar property of the matrix model (hypergeometric)
$\tau$-functions,
which is actually a combination of these two
complementary requirements.
\end{abstract}

\bigskip

\bigskip

\paragraph{Main formulas.} Recent studies of the two advanced subjects,
Hurwitz partition functions in \cite{AMMN2} and
rainbow tensor models in \cite{rainbow},
led to a spectacular new discovery in the old field
of the one-matrix models, which goes far beyond
their well-established theory presented in \cite{UFN3}
(for alternative descriptions see also \cite{MaMo}).
They concern the Hermitian and rectangular complex matrix models
given by the extended partition functions, respectively
\be\label{HMM}
Z_N\{t\} ={\displaystyle{ \int_{N\times N} dM \exp \left(-\frac{\mu}{2}\,\Tr M^2
+ \sum_k t_k \,\Tr M^k\right)}\over \int_{N\times N} dM \exp \left(-\frac{\mu}{2}\Tr M^2 \right)}
\ee
where the integral runs over the $N\times N$ Hermitian matrices,
and
\be\label{CMM}
Z_{N_1,N_2}\{t\} ={\displaystyle{ \int_{N_1\times N_2} d^2M
\exp \left(-\mu\,\Tr M M^\dagger + \sum_k t_k\, \Tr (MM^\dagger)^k\right)}
\over \int_{N_1\times N_2} d^2M\exp \left(-\mu\Tr M M^\dagger \right)}
\ee
where the integral is over {\it rectangular} $N_1\times N_2$ complex matrices.
With this choice of action, the second model has a symmetry $U(N_1)\times U(N_2)$,
much bigger than just a single $U(N)$ in the case of Hermitian model.

\bigskip

In fact, what have been known so far about these models is that
\begin{itemize}
\item $Z_N\{t\}$ and $Z_{N,N}\{t\}$ are actually the $\tau$-functions of the Toda chain, i.e. satisfy the
infinite system of bilinear Hirota equations \cite{MMint}
\item $Z_N\{t\}$ and $Z_{N_1,N_2}\{t\}$ satisfy the infinite system of recursive Virasoro constraints (see \cite{MMVir} and \cite{CMM,AMM3} respectively)
\be
\left(-\mu\frac{\partial}{\partial t_{n+2}}
+\sum kt_k \frac{\partial}{\partial t_{k+n}}
+ \sum_{a=1}^{n-1} \frac{\partial^2}{\partial t_a\partial t_{n-a}}
+2N\frac{\partial}{\partial t_n}
+ N^2\delta_{n,0}
\right)  Z_N\{t\}=0,
\ \ \ \ n\geq -1
\label{VirH}
\ee
\be
\left(-\mu\frac{\partial}{\partial { t_{n+1}}} +
\sum_k k{ t_k} \frac{\partial}{\partial { t_{k+n}}} +
\sum_{a=1}^{n-1} \frac{\partial^2}{\partial { t_a}\partial { t_{n-a}}}
+ (N_1+N_2)\cdot\,(1-\delta_{n,0})\,\frac{\partial}{\partial { t_{n}}}
+ N_1N_2\cdot \,\delta_{n,0} \right)
Z_{N_1,N_2}\{t\} = 0
\ \ \ \ n\geq 0
\label{VirC}
\ee
\end{itemize}

\noindent
This was already a lot and made these "matrix model" $\tau$-functions the most
important new special functions for the modern epoch of string theory calculations
\cite{UFN2,UFN3}.
The comprehensive theory of genus expansion,
the AMM/EO topological recursion \cite{AMM/EO},
a non-perturbative treatment \cite{AMM1} of the Dijkgraaf-Vafa phases \cite{DV,CM,DVmore,AMM1,AMM2,M,MMZ}
and the check-operators \cite{AMM2}
provided a dream-like standards for non-perturbative calculations
and were already applied in a variety of topics
ranging from knot theory \cite{Dijrec,MaMoknot}
to wall-crossing \cite{GMM} and AGT relations \cite{AGT}.

\bigskip

However, a long-standing problem overshadowed all these successes:
they resulted from an artful combination and interplay of the two features: the
Virasoro constraints and integrability.
These two are intimately related: from the early days it was realized \cite{FKN,MMint}
that integrability makes the entire infinite set of Virasoro relations into
a corollary of the lowest one named "string equation".
From this perspective, integrability describes the partition function in terms
of the Grassmannian, and the string equation picks up a concrete point in the Grassmannian.
This was a very big conceptual achievement, still its power is limited:
to get an answer to any concrete question, which in quantum field theory is
a calculation of the concrete correlation function, one still needs to do a complicated work
growing with the complexity of the correlator.

Worse of all, while the theory was nicely developing in the direction of
non-perturbative calculus, in the background of non-vanishing coupling
$t_k=T_k$, the Gaussian averages at $t=0$ remained untamed.
One sometimes manages to artfully deduce a particular formula for a class
of correlators like the famous Harer-Zagier answer \cite{HZ,AMM1}
for the arbitrary Gaussian average
\be
\sum_{k=0}^\infty \frac{z^{k}}{(2k-1)!!}\Big<{\rm Tr}_{_{N\times N}} M^{2k}\Big>
= \frac{\mu}{2z}\left(\left(\frac{\mu+z}{\mu-z}\right)^N-1\right)
\label{HZ1pH}
\ee
for any single-trace operator in the Hermitian model
(see \cite{ShaHZ} on generalizations to two- and three-trace case based partly on
related achievements in \cite{Ok}).
Nevertheless, it remains more a piece of art than a theory capable of producing answers
to arbitrary questions.

A slightly better result was a $W$-representation \cite{ShaW}, where the partition function
is represented as an action of integrability-preserving operator $\hat W$ on a trivial
$\tau$-function, e.g
\be\label{HMMW}
Z_N\{t\} = e^{\frac{1}{2\mu}\hat W\{t\}}\cdot e^{Nt_0}
\ee
with
\be
\hat W_{2} = \sum_{a,b} \left(
abt_at_b\frac{\partial}{\partial t_{a+b-2}} +
(a+b+2)t_{a+b+2} \frac{\partial^2}{\partial t_a\partial t_b}\right)
\label{cjoH}
\ee
and
\be\label{CMMW}
Z_{N_1\times N_2}\{t\} =
\exp\left\{{1\over\mu}\Big(N_1N_2t_1+(N_1+N_2)\hat L_{1} + \hat W_{1}\Big)\right\}\cdot 1
\ee
with
\be
\hat L_1 =\sum_m (m+1)t_{m+1}\frac{\p}{\p t_m},  \\
\hat W_1 =\sum_{a,b} abt_at_b \frac{\p}{\p t_{a+b-1}} + (a+b+1) t_{a+b+1}\frac{\p^2}{\p t_a\p t_b}
\ee
Since the operators $\hat W$, $\hat L$ have non-vanishing grading $2$ and $1$,
each correlator of a given
grading appears in just {\it one} term of the expansion of the exponential
and thus becomes a solution to a finite-dimensional problem: that of iterative action
of finite number of $\hat W$'s on $e^{Nt_0}$ and of $\hat W$ and $\hat L$'s on unity.
Still, no explicit formula for a generic correlator was so far obtained in this way.

However, as a byproduct of recent studies of far more complicated problems
in \cite{AMMN2} and \cite{rainbow},
we introduced a set of formulas, some of them being probably new,
which do provide a {\bf full solution} of the Hermitian and rectangular
complex models.
Since those papers were not focused on these particular results, they remained
not-so-well-noticed, while they certainly deserve a dedicated and absolutely clear
presentation.
Namely, the partition functions have very explicit character expansions
and arbitrary Gaussian correlators are represented by finite sums.
Formulas are much simpler for the rectangular complex model with a bigger symmetry (see earlier results in \cite{KPSS,Ramg}):
\be
\boxed{
Z_{N_1\times N_2}\{t\} = \sum_{R} \mu^{-|R|}\frac{D_R(N_1)\,D_R(N_2)}{d_R} \cdot \chi_R\{t\}
}
\label{charexpC}
\ee
\be\label{corrsC}
\boxed{
{\cal O}_{\Lambda} = \left<\prod_{i=1}^{l_\Delta} \Tr (M M^\dagger)^{l_i}\right>_G ={1\over\mu^{|\Lambda|}}
\sum_{R\, \vdash |\Lambda|}  \frac{D_R(N_1)\,D_R(N_2)}{d_R}  \cdot \psi_R(\Lambda)
}
\ee
where $\Lambda = \{l_1\geq l_2\ldots\geq l_\Delta >0\}$ and $R$ are the Young diagrams
of the given size (number of boxes) $|\lambda| = \sum_i l_i$,
and $D_R(N)$, $\chi_R\{t\}$, $\psi_R(\Lambda)$ and  $d_R$ are respectively the  dimension
of representation $R$ for the linear group $GL(N)$, the linear character (Schur polynomial),
the symmetric group character and the dimension of representation $R$ of the symmetric group $S_{|R|}$ divided by $|R|!$,
$D_R(N) = \chi_R\{t_n = N/n\}$, $d_R = \chi_R\{t_n=\delta_{n,1}\}$ \cite{Fulton}. Further we will also need a factor
$z_{\Lambda}$, which is the standard symmetric factor of the Young diagram (order of the automorphism) \cite{Fulton}.

For the Hermitian model, the structure of the formulas is the same,
but the Young diagrams are of even dimension (number of boxes) and additional
weight factors emerge made from differently-normalized symmetric characters
$\varphi_R(\Delta) = \psi_R(\Delta)\cdot d_R^{-1}\cdot z_\Lambda^{-1}$ from \cite{MMN1} (cf. also with a Fourier expansion of \cite[Eq.(2.18)]{MKR} in terms of characters):
\be
\boxed{
Z_{N }\{t\} = \sum_{R \ {\rm of\ even\ size}}
\mu^{-|R|}\varphi_R\Big(\underbrace{[2,\ldots,2]}_{|R|/2}\Big)\cdot D_R(N) \cdot \chi_R\{t\}
}
\label{charexpH}
\ee
\be
\boxed{
{\cal O}_{\Lambda} = \left<\prod_{i=1}^{l_\Delta} \Tr M^{l_i}\right>_G =
{1\over\mu^{|\Lambda|}}\sum_{R\, \vdash |\Lambda|}  \varphi_R\Big(\underbrace{[2,\ldots,2]}_{|R|/2}\Big)\cdot D_R(N) \cdot \psi_R(\Lambda)
}
\label{corrsH}
\ee
All ingredients in these formulas are well known objects from the
basic representation theory, the least trivial of them, $\psi_R(\Lambda)$
are called by the MAPLE command $Chi(R,\Lambda)$ in the {\it combinat} package.
The strangely-looking factor in (\ref{corrsH}) is actually
\be
\varphi_R\Big(\underbrace{[2,\ldots,2]}_{|R|/2}\Big) = \frac{1}{d_R}
\chi_R\left\{t_{n} = \frac{1}{2}\delta_{n,2}\right\}
\ee

\bigskip

{\bf The four formulas (\ref{charexpC})-(\ref{corrsH}) provide
a complete solution to the models,
much more powerful than just integrability or Virasoro constraints.}

\bigskip

These formulas (\ref{charexpC}), (\ref{charexpH}) once again emphasize
a relative simplicity of the complex
model of \cite{CMM} with respect to the more familiar Hermitian one.
For example, the very first check of integrability, the simplest Pl\"ucker
relation
\be
w_{[2,2]}w_{[0]} - w_{[2,1]}w_{[1]} + w_{[2]}w_{[1,1]} = 0
\ee
between the coefficients $w_R$ in the character expansion
of the KP tau-function $\tau\{t\} = \sum_R w_R\chi_R\{t\}$ in the case of
(\ref{charexpH}) is just an obvious identity between the dimensions
{\footnotesize
$$
\frac{N_1^2(N_1^2-1) \,N_2(N_2^2-1)}{12} -
\frac{N_1(N_1^2-1)\,N_2^2(N_2-1)}{3}\cdot N_1N_2
+ \frac{N_1(N_1+1)\,N_2(N_2+1)}{2}\cdot \frac{N_1(N_1-1)\,N_2(N_2-1)}{2}
= 0
$$
}
while for (\ref{charexpH}) it  turns into
$$
3\cdot \frac{N^2(N-1)^2}{12} - 0 - \frac{N(N+1)}{2}\cdot \frac{N(N-1)}{2} = 0
$$
which is  algebraically even simpler, but conceptually far less obvious,
because of the need for the strange coefficients $3,0,-1$,
which as one realizes are made from the characters $\varphi$.

It is amusing that despite (\ref{charexpC})-(\ref{corrsH}) provide
{\bf exact and exhaustive solutions} to the two models,
none of their celebrated properties \cite{UFN3} are obvious from these formulas:
neither integrability \cite{MMint,AMM3} nor Virasoro relations \cite{MMVir,CMM,AMM3},
neither $W$-representations \cite{ShaW,AMMN2}, nor Harer-Zagier formulas \cite{HZ,AMM1,ShaHZ}:
to be revealed, they all still require rather artful arguments and calculations.
This adds to the mysteries of matrix models and emphasizes that,
despite being {\it perturbatively} solved now, they will remain a source
of puzzles and inspiration for further theoretical investigations.

\bigskip

One of the claims in \cite{AMMN2} was that the extended partition functions of the Hermitian
and complex matrix models belong to a peculiar family of
the hypergeometric $\tau$-functions introduced in \cite{GKM2,OS},
which are the sums
\be
Z_{(k,m)}\Big\{\mu,N_1,\ldots,N_m\,|\,t^{(i)}\Big\} =
\sum_R \mu^{-|R|} d_R^{2-k-m}  \left( \prod_{i=1}^k \chi_R\{t^{(i)}\}\right)
\left( \prod_{i=1}^m D_R(N_i)\right)
\label{Zotau}
\ee
with $k\leq 2$.
We refer to \cite{AMMN2} for a thorough study of the entire family,
while here we concentrate on our two models
which we now identify as
\be
{\rm Rectangular\ complex\ model} && Z_{N_1\times N_2}\{t\}=Z_{(1,2)}\Big\{\mu,N_1,N_2|t_k\Big\}=Z_{(2,2)}\Big\{\mu,N\,\Big|\,\bar t_k=\delta_{k,1},\ t_k\Big\}
\nn \\ \nn \\
{\rm Hermitian\ model} && Z_N\{t\}=Z_{(2,1)}\left\{\mu,N\,\Big|\,\bar t_k={1\over 2}\delta_{k,2},\  t_k\right\}
\ee
and add some more technical details,
helpful for making a bridge between the new and old approaches.

\paragraph{Rectangular complex matrix model of \cite{CMM,AMM3} (RCM).}
Due to its higher symmetry,
this kind of a complex matrix model (\ref{CMM})
is simpler than the Hermitian model, despite dealing with rectangular matrices.
In fact, one can immediately establish its representation
in the form (\ref{charexpC})
by noticing that the Virasoro constraints (\ref{VirC}),
which unambiguously determine the partition function,
coincide with those for the partition function $Z_{(1,2)}$ (\ref{Zotau}),
which has the representation (\ref{charexpC})
(note that the matrix model in \cite{AMMN2} describing this partition function,
though being a complex matrix model, is looking a bit different):
\be
Z_{N_1,N_2}\{t\}=Z_{(1,2)}\Big\{\mu,N_1,N_2\,\Big|\,t_k\Big\}
\ee
Hence, one can immediately read off the results of \cite{AMMN2},
in particular, get the $W$-representation (\ref{CMMW}).
On the other hand, the character expansion representation (\ref{charexpC})
means that the partition function (\ref{CMM})
is a hypergeometric KP $\tau$-function \cite{GKM2,OS},
i.e. is the $\tau$-function of the form
\be
\tau=\sum_R d_R\chi_R\{t\}\cdot \left(\prod_{i,j\in R} f(i-j)\right)
\label{taf}
\ee
In the concrete case of RCM, the function
$f(x)=(N_1+x)(N_2+x)$, since
\be
{D_R(N)\over d_R}=\prod_{i,j\in R}(N+i-j)=\prod_i{(\lambda_i+N-i)!\over (N-i)!}
\ee

\paragraph{On Toda deformations of rectangular complex model.}
The hypergeometric $\tau$-function of the KP hierarchy,
which depends on {\it one} infinite set of times,
can be immediately continued to the Toda lattice $\tau$-function depending
on {\it two} infinite sets of times $\{t\}$, $\{\bar t\}$ and one discrete zero-time $n$:
\be
\tau_n=\sum_R \chi_R\{t\}\chi_R\{\bar t\} \cdot \left(\prod_{i,j\in R} f(n+i-j)\right)
\label{taTo}
\ee
According to (\ref{Zotau}),
what one gets in this way, is the partition function $Z_{(2,2)}$.

This continuation to the Toda lattice is, however, not unique.
Indeed, let us consider the particular case of RCM with $N_1=N_2=N$,
i.e. the {\it square} complex model of \cite{CMM,AMMN2} which still has
an enhanced symmetry $U(N)\times U(N)$
(i.e. belongs to the {\it rainbow} class in the modern terminology of \cite{rainbow}).
Then, from \cite{CMM,AMMN2} we know that
 the partition function is a Toda chain $\tau$-function
with the role of the discrete time  played by the size of the complex matrix $N$.
This can be seen, for instance, from its determinant representation:
\be
Z_{N,N}\{t\}=\det_{0\le i,j\le N}C_{i+j},\ \ \ \ \ \ \ \
C_k=\int_0^\infty x^kdx\  \exp\Big(-\mu x+\sum_kt_kx^k\Big)
\ee
However, the Toda chain $\tau$-function is a special solution
to the Toda lattice that depends only on the {\it difference} of the two sets of Toda times.
This is not the case for $Z_{(2,2)}$, hence
(\ref{taTo}) is a different continuation to the Toda lattice hierarchy,
not leading to the Toda chain.
In other words, there is another representation of (\ref{CMM})
in addition to (\ref{taf}), i.e. to (\ref{taTo}) evaluated at $\bar t_k=\delta_{k,1}$.
It should look like
\be
Z_n\{t-\bar t\}=\sum_{R,R'}c_{RR'}(n)\cdot \chi_R\{t\}\chi_R\{\bar t\}
\ee
with some peculiar non-diagonal coefficients $c_{RR'}(n)$, adjusted to make the r.h.s.
a function of only the differences $\{t_k-\bar t_k\}$.

In other words, there are two ways to make a Toda-lattice  $\tau$-function
from the function (\ref{CMM}):
the one available at any $N_1$ and $N_2$ is to {\it extend} it:
\be
\sum_R \frac{D_R(N_1)D_R(N_2)}{d_R}\cdot \chi_R\{t\} \ \longrightarrow \
\sum_R \frac{D_R(N_1)D_R(N_2)}{d_R^2}\cdot \chi_R\{t\}\chi_R\{\bar t\}
\ee
and another one working only at $N_2=N_1$, but requiring {\it no} deformations,
is just a change of arguments  $t\longrightarrow t-\bar t$:
\be
{\rm as\ it\ is,} \ \ \ \
Z_{N\times N}\{t-\bar t\} \ \ \ \
{\rm is\ {\it already}\ a\ Toda\ chain\ \tau-function}
\ee
with $N$ playing the role of the discrete Toda time.

\paragraph{Gaussian matrix model.}

The case of the Hermitian matrices looks   more complicated, though the character decomposition (\ref{charexpH})
allows one to identify it with $Z_{(2,1)}$
(which in \cite{AMMN2} was actually associated with a different matrix model):
\be
Z_{N }\{t\} = \sum_{R \ {\rm of\ even\ size}}
\frac{1}{\mu^{|R|}}\,
\varphi_R\Big(\underbrace{[2,\ldots,2]}_{|R|/2}\Big)\cdot D_R(N) \cdot \chi_R\{t\}
= \nn \\
= \sum_{R }
\frac{1}{\mu^{|R|}} \frac{D_R(N)}{d_R}
\cdot \chi_R\{t\}\chi_R\left\{\bar t_k= {1\over 2}\delta_{k,2}\right\}\
= \ Z_{(2,1)}\left\{\mu,N\,\Big|\,\bar t_k={1\over 2}\delta_{k,2},\  t_k\right\}
\ee
since it follows from
\be
\chi_R\{t\}=\sum_{\Lambda\vdash |R|}d_R\varphi_R(\Lambda)p_\Lambda
\ee
that
\be
\chi_R\left\{\bar t_k={1\over 2}\delta_{k,2}\right\}=\left\{
\begin{array}{lr}
d_R\varphi_R\Big(\underbrace{[2,\ldots,2]}_{|R|/2}\Big)&\hbox{for even }|R|\\
0&\hbox{for odd }|R|
\end{array}
\right.
\ee
Thus, it is again a KP $\tau$-function of the hypergeometric type.
And again, like the complex model case, this KP $\tau$-function
can be extended to the Toda lattice case in a different way
from switching on another set of times in $Z_{(2,1)}$:
namely,  by considering $Z_N\{t\}$ as a $\tau$-function
of the Toda chain with $N$ being the discrete Toda time,
\be
Z_{N}\{t\}=\det_{0\le i,j\le N}H_{i+j},\ \ \ \ \ \ \ \
H_k=\int_{-\infty}^\infty x^k dx  \ \exp\Big(-{\mu x^2\over 2}+\sum_kt_kx^k\Big)
\ee

\paragraph{Correlators from character expansion.}
The simple combinatorics needed to go from
(\ref{charexpC}) to (\ref{corrsC}) and back
involves the Cauchy formula \cite{Fulton}
\be\label{ex1}
\left<e^{\sum_kt_k\Tr M^k}\right>=\sum_\Lambda {1\over z_\Lambda}p_\Lambda\cdot {\cal O}_\Lambda,\ \ \ \ \ \ p_\Lambda\equiv \prod_i l_it_{l_i}
\ee
for the correlators ${\cal O}_\Lambda$
and the identity
\be
p_\Lambda=\sum_{R\vdash |\Lambda|}\psi_R(\Lambda)\chi_R\{t\}
\ee
for monomial functions of times. Then
\be
Z_{N_1\times N_2}\{t\} =\left<e^{\sum_kt_k\Tr (MM^\dagger)^k}\right>=
\sum_{R\vdash |\Lambda|,\Lambda} {1\over z_\Lambda}\psi_R(\Lambda)\cdot {\cal O}_\Lambda\cdot
\chi_R\{t\}
\ee
Thus, if we denote the coefficient in the character expansion of the partition
function by $\mathfrak{N}_R$,
\be
\boxed{
Z=\sum_R\mathfrak{N}_R\cdot\chi_R\{t\}
}
\ee
we see that it is expressed through the correlators:
\be
\mathfrak{N}_R=\sum_\Lambda {1\over z_\Lambda}\psi_R(\Lambda)\cdot {\cal O}_\Lambda
\ee
Using the orthogonality relation
\be
\sum_R {1\over z_\Delta}\psi_R(\Delta)\psi_{R}(\Delta ')=\delta_{\Delta\Delta '}
\ee
one can invert this relation:
\be\label{ex2}
\boxed{
{\cal O}_\Lambda=\sum_R\mathfrak{N}_R\cdot \psi_R(\Lambda)
}
\ee
i.e. ${\cal O}_\Lambda$ and $\mathfrak{N}_R$ are connected
by a transformation with the kernel $\psi_R(\Lambda)$.

\bigskip

With this general knowledge, one can easily convert the character expansion (\ref{charexpC})
for the rectangular complex model
into  (\ref{corrsC}) and back.
In the same way, with the only substitution $MM^\dagger\to M$,
one can relate (\ref{corrsH}) with the character expansion (\ref{charexpH}).

\paragraph{Summary.}
In this paper, we explicitly described {\bf complete solution} in the perturbative
phase to the most famous Hermitian matrix model: this means that we provided a
concise and clear {\bf answer} for arbitrary Gaussian correlator,
a nice complement to  the celebrated Harer-Zagier formula for a particular family\footnote{
One could expect that, for any given correlator, obtaining an explicit answer using a formulae of the Harer-Zagier type, which involves
expanding some generating functions would be computationally less costly than summing over Young diagrams of size n with a weight involving characters. However, in fact, summation over diagrams is actually not so difficult with Mathematica and MAPLE, which have built-in symmetric group characters. More important, however, is that Harer-Zagier type formulas for generic correlators are still unknown and constitute one of the challenging problems to be solved in matrix model theory.}.
Moreover, there is a theory, where such formulas are even simpler:
this is the rectangular complex matrix model {\it with enhanced symmetry},
strongly advertised in \cite{CMM}, but unjustifiably underestimated since then
(fortunately, not by all, see, for example, \cite{Ramg}).
The recent papers \cite{AMMN2} and \cite{rainbow} describe far going applications and generalizations
of these results, but clearly there can and will be many other.
Also the theoretical understanding and even various derivations of these formulas
constitute a challenge, which can have strong impact.
All these questions are, however, beyond the scope of this letter, which is targeted at
the maximally concise and clear formulation of the very important solvability
statement.

\section*{Acknowledgements}

This work was funded by the Russian Science Foundation (Grant No.16-12-10344).

\end{document}